# DEFINITE SOLUTION OF THE TWO CAPACITORS PARADOX (AND TWO WATER BUCKET PARADOX)

Vladan Panković

Department of Physics, Faculty of Sciences, 21000 Novi Sad, Trg Dositeja Obradovića 4., Serbia, vladan.pankovic@df.uns.ac.rs

## Abstract

In this work we suggest very simple solution of the two capacitors paradox in the completely ideal (without any electrical resistance or inductivity) electrical circuit. Namely, it is shown that electrical field energy loss corresponds to works done by electrical fields of both capacitors by movement of the electrical charge. It is all and nothing more (some dissipative processes, e.g. Joule heating and electromagnetic wave emission effects) is necessary. Additionally, we shortly demonstrate a simple mechanical analogy of mentioned paradox and its solution. Concretely we consider two water buckets connected by a valve where is a seeming loss of the gravitational potential energy that can be explained by the work in the gravitational field.

As it is well-known in many remarkable textbooks and articles referring on the basic electrical and electro-dynamical principles and their applications [1]-[7] two capacitors problem or paradox is formulated and considered. This paradox, simply speaking, states the following. By charging of the second, initially non-charged, by the first, initially charged, capacitor in an ideal (without any electrical resistance and inductivity) electrical circuit there is loss of one half of the initial energy of the electrical field. It is well-known too that there are many different attempts of the solution of mentioned paradox [4]-[7]. Generally speaking these solutions suppose that electrical circuit cannot be ideal, even not approximately, or, that some small (residual) electrical resistances or/and inductivities must exist. Then given electrical field energy loss can be explained by dissipative, thermal processes (Joule heating) or electro-magnetic waves emission. However, all this needs very complex theoretical formalism in distinction to theoretically extremely simple formulation of the paradox itself.

In this work we shall suggest very simple solution of the two capacitors paradox in the completely ideal (without any electrical resistance or inductivity) electrical circuit. Namely, it is shown that electrical field energy loss corresponds to works done by electrical fields of both capacitors by movement of the electrical charge. It is all and nothing more (some dissipative processes, e.g. Joule heating and electromagnetic wave emission effects) is necessary. Additionally, we shall shortly demonstrate a simple mechanical analogy of mentioned paradox and its solution. Concretely we shall consider two water buckets connected by a valve where is a seeming loss of the gravitational potential energy that can be explained by the work in the gravitational field.

As it is well-known [1]-[7] two capacitors paradox can be formulated in the following way.

Consider a simple electrical circuit that holds only one switch and two identical capacitors with the same capacitance C, without any resistance and inductivity. Initially, switch is in the state OFF so that electrical circuit is open. Then first capacitor is charged by electrical charge Q and holds voltage $V = QC$, while second capacitor is without charge and voltage. Energy of the electric field within the first capacitor equals, as it is well-known,

(1) $\quad E_{1in} = 1/2 CV^2 = Q^2/2C$

while electrical field (energy) within second capacitor does not exist at all. For this reason total energy of the electrical fields within both capacitors $E_{in}$ equals $E_{1in}$ (1) too.

But when switch turns out in the state ON electrical circuit becomes closed and during a very small time interval both capacitors become oppositely charged with electrical charges Q/2 and −Q/2 and hold opposite voltages V/2 and −V/2. In other words, during this small time interval charge Q/2 turns out from the first on the second capacitor. Then, energies of the electric field within the first and second capacitor are identical. Given energies equal, as it is well-known,

(2) $\quad E_{1fin} = E_{2\,fin} = 1/2 C(V/2)^2 = (Q/2)^2/2C = 1/4\, Q^2/2C$

so that total energy of the electrical fields within both capacitors equals

(3) $\quad E_{fin} = E_{1fin} + E_{2fin} = 1/2\, Q^2/2C = 1/2\, E_{1in} = 1/2\, E_{in}$ .

In this way there is the following, seemingly paradoxical, energy loss

(4) $\quad \Delta E = E_{fin} - E_{in} = -1/2\, E_{in} = -Q^2/(4C)$

or energy difference between the final and initial state of given electrical circuit.

However, it can be observed that, according to well-known definitions and (1), (2), the following differences

(5) $\quad A_1 = E_{1fin} - E_{1in} = -3/4\, E_{in}$
(6) $\quad A_2 = E_{2fin} - E_{2in} = 1/4\, E_{in}$

represent works done by the electrical field of the first and second capacitors respectively by movement of the electrical charges, Then

(7) $\quad A = A_1 + A_2 = -1/2\, E_{in}$

represents total work of the both fields by movement of the electrical charges. Obviously this work (7) is identical to energy difference (4). In other words, in full accordance with basic electro-dynamical concepts, decrease of the energy of the electrical field corresponds to the work done by the electrical charges movement using mentioned field. It is all and nothing more (some dissipative processes, e.g. Joule heating and electromagnetic wave emission effects) is necessary. (More precisely we shall consider that really existing resistance and inductivities yield only high order corrections which here can be neglected.)

In this way we obtain very simple and reasonable solution of the two capacitors paradox in the completely ideal (without any electrical resistance or inductivity) electrical circuit.

Now we shall shortly demonstrate a simple mechanical analogy of mentioned paradox and its solution.

Consider two identical, cylindrical buckets with high H. These buckets are mutually connected at their bottoms using a small horizontal valve.

Initially valve is in OFF state and first bucket is completely filled with the water while second bucket is completely empty. If M represents the mass of the water in the completely filled bucket, initial total potential gravitational energy of the water in the first bucket can be expressed as

(8) $\quad E_{1in} = MgH/2$

where g represents Earth gravity acceleration, while H/2 represents position of the bucket water mass center in respect to the bottom. Since initially in the second bucket there is no water at all initial total potential gravitational energy of the water in the first bucket does not exist at all. In this way total potential energy of the water in both buckets $E_{in}$ equals $E_{1in}$ (9) too.

Further, we can set valve in ON state so that water begins to flow from the first in the second bucket till the final moment when both buckets are half full. More precisely in the final moment level of the water in any bucket equals H/2 and corresponding mass equals M/2. Then final total potential gravitational energy of the water in any bucket equals

(9)     $E_{1fin} = E_{2fin} = (M/2) \, g \, (H/2) \, /2$

where $(H/2)/2 = H/4$ represents position of the bucket water mass center in respect to the bottom. In this way final total potential gravitational energy of the water within both buckets equals

(10)     $E_{fin} = E_{1fin} + E_{2fin} = MgH/4 = E_{1in}/2$     .

what is two time smaller than corresponding initial total potential gravitational energy of the water in both buckets.

In this way there is the following, seemingly paradoxical, total potential gravitational energy loss

(11)     $\Delta E = E_{fin} - E_{in} = -1/2 \, E_{in}$

or energy difference between the final and initial state of the water in both buckets.

Thus, we obtained a complete mechanical analogy with two capacitor paradox. Solution of mentioned mechanical paradox can be obtained in analogous way too.

Namely, it can be observed that, according to well-known definitions and (8), (8), the following differences

(12)     $A_1 = E_{1fin} - E_{1in} = -3/4 \, E_{in}$
(13)     $A_2 = E_{2fin} - E_{2in} = 1/4 \, E_{in}$

represent works done by the Earth gravitational field in the first and second bucket respectively by movement of the water. Then

(14)     $A = A_1 + A_2 = -1/2 \, E_{in}$

represents total work of the Earth gravitational field by movement of the water. Obviously this work (14) is identical to energy difference (11). In other words, in full accordance with basic mechanical concepts, decrease of the energy of the gravitational field corresponds to the work done by the water mass movement using mentioned field. It is all and nothing more (some dissipative processes, e.g. friction between water and bucket walls) is necessary. (More precisely we shall consider that really existing mentioned dissipative processes yield only high order corrections which here can be neglected.)

In conclusion, the following can be shortly repeated and pointed out. In this work we suggest very simple solution of the two capacitors paradox in the completely ideal (without any electrical resistance or inductivity) electrical circuit. Namely, it is shown that electrical field energy loss corresponds to works done by electrical fields of both
capacitors by movement of the electrical charge. It is all and nothing more (some dissipative processes, e.g. Joule heating and electromagnetic wave emission effects) is necessary. Additionally, we shortly demonstrate a simple mechanical analogy of mentioned paradox and its solution. Concretely we consider two water buckets connected by a valve where is a seeming loss of the gravitational potential energy that can be explained by the work in the gravitational field.

.